# On quantum critical destruction of the pseudogap phase

N. Kristoffel

Institute of Physics, University of Tartu, W. Ostwald Str. 1, 50411 Tartu, Estonia

*Recent data on the critical events accompanying the quenching of the pseudogap (PG) phase in hole-doped cuprates are analysed. An original multiband model with the spectrum influenced by doping has been used. Two critical dopings are expected in evolving of the PG to destruction. The first of them corresponds to metallization (antinodal) of the sample. Extended doping leads to a second critical point where the PG is lost. In between the restoration of the full Fermi surface proceeds. Band overlaps besides the location of the chemical potential ($\mu$) are decisive.*

There are specific gapped low-energy excitations observed in both normal and superconducting states of cuprates. This has delivered the elaboration of the concept of the pseudogap (PG) [1-3]. The process of the formation of the doped electronic compound from basic charge-transfer (Mott) antiferromagnetic system remains under discussion. An enormous amount of investigations have analysed this mystery from different points of view. Various coupled electron and lattice subsystem distortions (like CDW) have been attached [4]. However, it has been shown that such factors lead to independent phases [5,6].

A recent accomplishment has been the demonstration of the existence of well-formed Fermi surfaces (or segments) in cuprate superconductors by observation of connected magnetic quantum oscillations [7,8]. Because the PG is an electronic excitation, its changes must be reflected in Fermi surface distributions. There is the process of the PG vanishing on the way from energetically exposed low-doping region to the scale comparable with superconducting excitations. It ends in the middle of the energy-doping phase diagram. Disappearance of the PG is usually ascribed to the presence of a quantum critical point. In this region elevated superconducting transition temperatures seem to be supported. One connects this circumstance with the liberation of additional electron states in favour of taking part in the superconductivity mechanism. These states have been hidden in the PG existing at weaker dopings. The material is expected to show accordingly an insulator to metal transition (crossover) [9,10] in the spectral region where the PG has resided. The whole development of the PG wears evidently basic information on the inner functionality of the superconductor. In the critical region the energy scales of the PG and superconducting gap are transformed to be comparable. Interference of the PG and of the superconductivity can be expected.

There have been difficulties in identifying the PG and the superconducting gap. The reason has been seemingly in ignorance of function of several bands in cuprate superconductivity and spectral windows representing various momentum space regions. Taking account [11] of these circumstances there can be realized a situation where e.g. a superconducting gap can be observed in the developed PG phase. These two gaps stem then from different bands. Besides the usual ARPES spectroscopic methods a number of others allow to follow the doping phase diagram of cuprates. Especially it is of interest to look on the carrier concentration during the passage of the PG to the critical region where it disappears. Remarkable novel results have been fixed on this passage [5,6]. Parallel investigation by Hall effect and of sample resistivity has essentially precised what happens with the charge carriers' number during the evolution of the PG state to its loss. Sudden changes in the carrier concentration near the critical point at the end of the PG have been observed [5,6]. In the underdoped region with the consummate PG the carrier concentration has been found to be equal to the doping concentration ($p$) i.e. $n = p$. The disappearance of the PG shifts it sharply to $n = 1 + p$. The point where the PG becomes zero exhibits also here its critical nature. It must be underlined that this transformation is between two metallic phases.

In the present note we point out that this transition process can be expected to be of extended nature. Two special (critical) dopings ($p_1 < p_0$) must be passed on the way to PG=0. In between of them the reconstruction of the FS is completed. This passage ends with the disappearance of the PG at $p_0$. Behind $p_0$ a quasi-normal (strange) electron liquid is formed. The carrier liquid becomes a mixture of all electron bands involved. The other critical point ($p_1$) corresponds to metallization of the sample in the spectral window characteristic for the PG bearing band. The superconducting gap and the PG show probably different symmetries at underdoping. Both mentioned critical points support distinct positioning of bands and location of $\mu$. These conclusions follow from our original model of cuprate interband superconductivity [11-14].

This model is essentially of multiband nature and exploits the electron spectrum substantially influenced by doping. There are new defect (polaronic) bands created by doping. Bottoms of these bands are lowered with doping. It is of central significance that doping changes the bands' overlap. The whole theoretical scheme uses our original definition of PG considering it as quasiparticle excitation [15]

$$\Delta_\alpha(PG) = \left\{ \left(\varepsilon_\alpha(\vec{k}) - \mu\right)^2_{min} + \Delta_\alpha \right\}^{\frac{1}{2}}.$$

Here $\varepsilon_\alpha(\vec{k})$ is the energy spectrum of the band on which the PG is formed, $\Delta_\alpha$ is the superconducting gap initiated in this band and $\mu$ is the chemical potential. For small difference $|\varepsilon_\alpha - \mu|$ the PG excitation can incorporate a superconducting contribution.

In multiband approaches there can be differences in positioning $\mu$ between the bands. It can be inside of the band $\alpha$, or out at them. In the first case $\left|\varepsilon_\alpha(\vec{k}) - \mu\right|_{min} \neq 0$ and the PG is present. It vanishes when $\left|\varepsilon_\alpha(\vec{k}) - \mu\right|_{min} = 0$ is reached, say, by doping. Then only $\Delta_\alpha$ plays a role and the $\alpha$-band excitation spectrum starts from zero frequency at $T = 0$ in the paraphase. The metallic configuration can be present also for overlapping with a partner band when $\left|\varepsilon_\alpha(\vec{k}) - \mu\right|_{min} \neq 0$ (i.e. PG phase) because the bottom of band, $\varepsilon_\alpha$, has not reached $\mu$.

The region between $p_1$ and $p_0$ must be suitable to extract the superconducting contribution to the PG. Theoretically this has been illustrated in [13]. In this manner the minimal value of its generic gap $|\varepsilon_\alpha - \mu|_{min}$ is decisive for the behaviour of the PG. It can be regulated by doping or other influences acting on $\varepsilon_\alpha$ and $\mu$ and their relation.

One utilizes the mentioned properties in the following discussion.

Accordingly the doping-stimulated process of the PG quenching must show two critical points $p_1$ and $p_0$. At $p_1$ the contribution to the metallization of the system appears from the PG-initiating band. The other one – $p_0$ – marks the end of the PG. The $\alpha$-band states contribute now to the building of the Fermi surface. Between $p_1$ and $p_0$ the metal behaves as a strange metal because of the preserved PG. This can explain the finding in [5,6] that the sudden change of carrier concentration appears between two metallic phases. The reconstruction of the Fermi surface connected with the PG-bearing band starts at $p_1$ and ends on $p_0$.

All these mentioned aspects enter our multiband model for a "typical" cuprate with interband mechanism under hole doping [11]. For the electron doping case, consult [12]. The bands' position dynamics including the changes in overlap and location $\mu$ of is illustrated in Fig. 1.

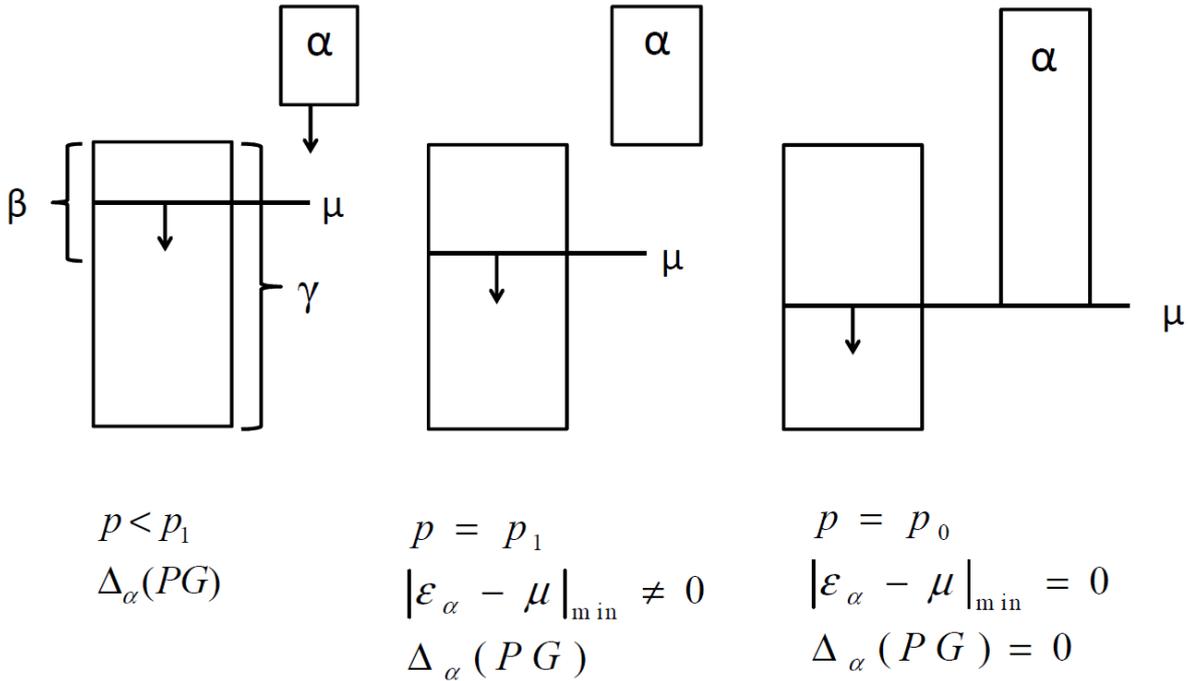

$p < p_1$     $p = p_1$     $p = p_0$
$\Delta_\alpha(PG)$     $|\varepsilon_\alpha - \mu|_{min} \neq 0$     $|\varepsilon_\alpha - \mu|_{min} = 0$
    $\Delta_\alpha(PG)$     $\Delta_\alpha(PG) = 0$

Fig. 1 Typical bands bearing $\mu$ configuration for $p < p_1$ (underdoping; pseudogap), $p = p_1$ (metallization, pseudogap) and $p = p_0$ (pseudogap closing). Note that the α-band bottom and $\mu$ are shifted down with doping.

It includes the itinerant band, predominantly of oxygen origin (nodal top), the doping created defect (polaronic) bands, $\alpha$ (antinodal bottom) and $\beta$ (nodal bottom). The latter are equally perturbed and formed at the cost of $\gamma$-band. In the problem under consideration the specific behaviour of $|\varepsilon_\alpha(\vec{k}) - \mu|_{min}$ is decisive.

At underdoping the rising $T_c$ is stimulated by the pair transfer interaction between $\beta$- and $\gamma$-bands. The hole antinodal pockets can be ascribed to $p < p_1$ situation. At $p_1$ the electron pockets start to contribute into conductivity and to determination of $\mu$. However, the PG remains present. When $p_0$ is reached the antinodal electron pockets give their full contribution to the FS. This kind of reorganisation has been pointed out in [5,6]. Behind $p_0$ the superconductivity is at last supressed. One of the reasons leading to this can be the destroyed $\beta - \gamma$ resonance.

As the conclusion we suppose that the quantum critical behaviour of the material near the ending of the PG can be expected to be extended including two phase diagram critical points. The detailed positioning of the change of the carrier concentration and the end of the PG on $p$-scale will be of interest.

**References**

1. T. Timusk, B. Statt, Rep. Prog. Phys. **62**, 61 (1999)
2. D. N. Basov and T. Timusk, Rev. Mod. Phys. **77**, 721 (2005)
3. G. Ghiringelli et al., Science **337**, 8121 (2012)
4. S. E. Sebastian and C. Proust, arXiv.1507.01315 (2015)
5. B. Vignolle et al., Nature **455**, 952 (2008)
6. F. Liberte et al., arXiv.1606.024491 (2016)
7. S. Badoux et al., Nature, **531**, 210 (2016)
8. W. Loram et al., J. Phys. Chem. Solids **59**, 209 (1998)
9. N. Kristoffel, P. Rubin and T. Örd, Int. J. Mod. Phys. B **22**, 5299 (2008)
10. N. Kristoffel and P. Rubin, Int. J. Mod. Phys. B **26**, 1250144 (2012)
11. N. Kristoffel and P. Rubin, Phys. Lett. A **374**, 79 (2009)
12. N. Kristoffel and P. Rubin, Solid State Commun. **122**, 265 (2002)
13. N. Kristoffel and P. Rubin, Phys. Lett. A **372**, 930 (2008)